\documentclass[ reprint,  amsmath,amssymb,  aps, aip,rsi,     ]{revtex4-1}%
\usepackage{graphicx}
\usepackage{dcolumn}
\usepackage{bm}
\usepackage{amsmath}
\usepackage{amsfonts}
\usepackage{amssymb}%
\providecommand{\U}[1]{\protect\rule{.1in}{.1in}}

\begin{document}

%\preprint{APS/123-QED}%

\title{Origin of Perpendicular Magnetic Anisotropy in Co$_x$Fe$_{3-x}$O$_{4+\delta}$ Thin Films \\
Studied by X-ray Magnetic Circular and Linear Dichroisms 
 }%

\author{Jun Okabayashi$^{1,*}$, Masaaki A. Tanaka$^2$, Masaya Morishita$^2$, Hideto Yanagihara$^3$, and Ko Mibu$^2$ }%

\affiliation{
$^1$Research Center for Spectrochemistry, The University of Tokyo, Bunkyo-ku, Tokyo 113-0033, Japan \\
$^2$Graduate School of Engineering, Nagoya Institute of Technology, Nagoya 466-8555, Japan \\
$^3$Institute of Applied Physics, University of Tsukuba, Tsukuba 305-8573, Japan 
}

%TCIMACRO{\TeXButton{\date{\today}}{\date{\today}}}%
%BeginExpansion
\date{\today}%
%EndExpansion
%

%TCIMACRO{\TeXButton{%
%\begin{abstract}%
%}{\begin{abstract}}}%
%BeginExpansion
\begin{abstract}%
%EndExpansion

We investigate the element-specific spin and orbital states and their roles on magnetic anisotropy in the Co-ferrite (Co$_x$Fe$_{3-x}$O$_{4+\delta}$ (001)) thin films which exhibit perpendicular magnetic anisotropy (PMA). The origin of PMA in the low $x$ region ($x$ $<$ 1) can be mainly explained by the large perpendicular orbital magnetic moments in the Co$^{2+}$ (3$d^7$) states detected by X-ray magnetic circular and linear dichroisms (XMCD/XMLD). The XMLD for a PMA film ($x=0.2$) with square hysteresis curve shows the oblate charge distribution in the Co$^{2+}$ site, which is consistent with the change in local nearest neighbor distance in Co detected by extended X-ray absorption fine structure analysis. Our finding reveals that the microscopic origin of PMA in Co-ferrite comes from the enhanced orbital magnetic moments along out-of-plane [001] direction through in-plane charge distribution by tensile strain, which adds the material functionalities in spinel ferrite thin films from the viewpoint of strain and orbital magnetic moments. 

\end{abstract}%

%TCIMACRO{\TeXButton{\maketitle}{\maketitle}}%
%BeginExpansion
\maketitle
%EndExpansion

%\section{\label{sec:level1}Introduction}

Perpendicular magnetic anisotropy (PMA) is one of the crucial issues in the spintronics research field from the viewpoints of thermal stability enhancement and high-density storage technology with the low current magnetization switching. Until now, tremendous efforts have been devoted to design the materials with large PMA to add the novel functionalities and external field controlling of magnetic properties \cite{ref1,ref2,ref3,ref4,ref5}. Among them, the magnetic oxides possess a number of attractive advantages for the spintronics devices because of their semiconducting or insulating behavior with high spin polarization. These magnetic oxides can be utilized as spin-filtering effects, resulting in high tunnel magnetoresistance with multiferroic properties by controlling the tensile or compressive strains in the thin films \cite{ref6,ref7,ref8}. 

Most famous spinel-type compounds for spintronics application might be Co-ferrites (CoFe$_2$O$_4$: CFO) epitaxial thin films because of their insulating advantages as spin-filtering, high-frequency performance, and magneto-strictive properties \cite{ref9,ref10,ref11,ref12,ref13,ref14,ref15,ref16}. Further, the interfacial proximity effect, emerging on films with non-magnetic heavy-metal elements which possess a large atomic spin-orbit interaction \cite{ref17,ref18,ref19}, also develops as the spin-orbitronics researches through the interfacial spin-orbit coupling. Recent development appends further functionality of large PMA in CFO in the order of 10$^6$ J/m$^3$ by tuning the Co compositions \cite{ref20,ref21,ref22,ref23}. Since CFO exhibits high magnetostriction coefficient ($\lambda_{100}=–590\times10^{-6}$), the strain from the substrates propagates into the CFO layer, which brings the local distortion around the transition-metal (TM) cation sites. In fact, the lattice mismatching between MgO (4.21 $\mathrm{\AA}$) and CFO (8.38 $\mathrm{\AA}$) forming almost the double period brings the epitaxial crystalline growth with a few percentages of tensile strain into CFO thin layer. Depending on the Co compositions, the lattice constants of CFO can be tuned systematically. Structural and magnetic properties in the CFO thin films can be controlled by the cation vacancies and anti-phase boundaries which are generated during the crystalline growth and suppress the saturation magnetization values through the antiparallel spin coupling \cite{ref24,ref25,ref26,ref27}. Historically, magnetic anisotropy in bulk CFO has been understood as an effect that the average of $\textless111\textgreater$ easy axes direction produces cubic anisotropy in Co$^{2+}$ \cite{ref28,ref29,ref30}. Recent theoretical and experimental investigations for CFO thin films suggest that the tensile strain into CFO layer produces the PMA \cite{ref31,ref32,ref33,ref34,ref35,ref36,ref37,ref38}. Squareness of out-of-plane hysteresis curves depends on the growth mode, such as molecular-beam epitaxy, pulsed-laser deposition (PLD), or sputter deposition. It is also reported that the PMA increases in off-stoichiometric CFO films with low Co compositions \cite{ref22,ref23}. Although the PMA in CFO is strongly demanded for spinel-type oxide spintronics and spin-orbitronics by controlling the orbital states \cite{ref39}, the microscopic origin of the PMA appearing in the low Co compositions of CFO has not been clarified yet. 

In order to understand the PMA in CFO, the element-specific orbital magnetic moments ($m_\mathrm{orb}$) and their orbital anisotropy have to be examined explicitly. In particular, $m_\mathrm{orb}$ is expected to be sensitive to the strain and local symmetry from the ligand fields around the TM ions. X-ray magnetic circular and linear dichroisms (XMCD / XMLD) with their magneto-optical sum rules are powerful tools to detect the element-specific spin and orbital states in each site with the charge asphericity \cite{ref40}. Especially, XMLD sum rules can probe the orbital anisotropy ($Q=3L_z^2-L^2$) in the notation of orbital angular momentum ($L$) and electric quadrupoles ($Q$). Furthermore, since the Fe $L$-edge spectra consist of the signals of three kinds of Fe$^{2+}$ or Fe$^{3+}$ states with octahedral $O_h$ and tetrahedral $T_d$ symmetries, the XMCD and XMLD magneto-optical sum rules for the integrals of Fe $L$-edge spectra cannot be adopted without separating each contribution. Site- and state-resolved analyses become a crucial role for understanding the origin of PMA. For the analysis of XMCD and XMLD spectra, the ligand-field multiplet (LFM) cluster-model simulations including the configuration interaction (CI) approach are used to determine the site-specific electronic structure parameters accompanied by spin and orbital magnetic moments and their anisotropies \cite{ref41}. Until now, the XMCD and XMLD for CFO with in-plane anisotropy have been extensively studied with angular dependence between incident polarized beam, magnetic field, and sample surface normal \cite{ref42,ref43,ref44,ref45}. However, the X-ray magnetic spectroscopy investigations for the PMA films with square magnetization–magnetic field ($M$-$H$) hysteresis shape in low Co compositions have not been pursued. We have proceeded the XMLD measurements for probing the orbital anisotropy in several PMA systems by using remanent magnetization states in the polar Kerr geometry \cite{ref46,ref47}. On the other hand, recent operando-XMCD technique enables manipulation of out-of-plane and in-plane $m_\mathrm{orb}$ by controlling strain in the magnetic multilayers, which is a microscopic origin of magneto-striction; that leads to the observation of an effect proposed as ''orbital-striction effect'' \cite{ref48}. The PMA in strained CFO can be also categorized as a novel orbital-controlled material system by using the orbital degeneracy in Co$^{2+}$ 3$d^7$ system in $O_h$ ligand field. Furthermore, extended x-ray absorption fine structure (EXAFS) analysis through the multiple-scattering process is also employed to determine the element-specific structural characterization. The precise determinations of spin and orbital states become a key solution to develop a strain-induced orbital physics and their applications using CFO. 

In this study, considering above research motivations, we aim to investigate the origin of PMA in CFO from the viewpoint of element-specific spin and orbital magnetic moments and their anisotropy using XMCD and XMLD with LFM calculations and local structure detection by EXAFS analysis.

The Co$_x$Fe$_{3-x}$O$_{4+\delta}$ (001) samples ($x$=0.2 and 0.6) were prepared by PLD on the MgO (001) substrates with the similar recipe of reference \cite{ref23}. The PLD was performed under the conditions of the substrate temperature of 300$^\circ$C, the background oxygen pressure of 6.0 Pa, and the deposition rate of 0.03 nm/s. A neodymium-doped yttrium-aluminum-garnet laser at the double frequency (532 nm) with the pulse width of 6 ns and the repetition rate of 30 Hz was used. The energy density of the laser beam was controlled to 1 J/cm$^2$ by an optical lens. The 13-nm-thick CFO (001) layer with 1-nm-thick Cu capping layer was deposited for various Co compositions ($x$). Excess oxygens accompanied by possible cation vacancies are described as $\delta$. The case of $x$=0.2 and 0.6 exhibits the PMA and in-plane anisotropy, respectively. The in-plane anisotropy in Co$_{0.6}$Fe$_{2.4}$O$_{4+\delta}$ is produced by the post annealing at 400$^{\circ}$C after the growth by releasing the strain.  The results of sample characterization by X-ray diffraction (XRD) and magnetization measurements are shown in Supplemental Material (Fig. S1) \cite{ref49SM}.

The XMCD and XMLD were performed at BL-7A and 16A in the Photon Factory at the High-Energy Accelerator Research Organization (KEK-PF). For the XMCD measurements, the photon helicity was fixed, and a magnetic field of $\pm$1.2 T was applied parallel to the incident polarized soft x-ray beam, to obtain signals defined as $\mu$+ and ${\mu}-$ spectra. The total electron yield mode was adopted, and all measurements were performed at room temperature. The X-ray absorption spectroscopy (XAS) and XMCD measurement geometries were set to normal incidence, so that the directions of photon helicity axis and the magnetic field were parallel and normal to the surface, enabling measurement of the absorption processes involving the normal components of the spin and orbital magnetic moments. In the XMLD measurements, the remanent states magnetized out-of-plane were adopted. For grazing incident measurements in XMLD, the tilting angle between incident beam and sample surface normal was kept at 60$^{\circ}$. The direction of the electric field ($\bf{E}$) of the incident linearly polarized synchrotron beam was tuned horizontally and vertically by undulator. We define the sign of XMLD by the subtraction of the ($\bf{M}{\parallel}{\bf{E}}$)$-$(${\bf{M}}{\perp}{\bf{E}}$) spectra with respect to the magnetization $\bf{M}$. The EXAFS measurements at Co $K$-edge were performed at BL-12C in KEK-PF using the florescence yield mode with a 19-element solid-state detector at room temperature. 

Figure 1 shows the XAS and XMCD of Co$_{0.2}$Fe$_{2.8}$O$_{4+\delta}$ for Fe and Co $L$-edges at the normal incidence setup. Spectra are normalized at each absorption edge. Because of the composition ratio of Co:Fe=1:14, the XAS intensities of Co are suppressed. XMCD intensity ratio to XAS is estimated to be 5$\%$ and 47$\%$ for Fe and Co $L_3$ edge, respectively. The raw data are displayed at the Supplemental Material (Fig. S2). XAS and XMCD line shapes for Fe $L$-edges show distinctive features due to the three kinds of Fe states (Fe$^{3+}$ in $O_h$, Fe$^{3+}$ in $T_d$, and Fe$^{2+}$ in $O_h$). For the Fe $L$-edges, although the difference in XAS is small, clear differential XMCD line shapes are detected. The Fe$^{3+}$ state with $T_d$ symmetry exhibits opposite sign, which is common for the spinel ferrite compounds. On the other hand, in the case of Fe$_3$O$_4$, the Fe$^{2+}$ component is more enhanced \cite{ref49}. Large XMCD signals in Co $L$-edge correspond to the saturated magnetized states. Within the orbital sum rule, the large $m_\mathrm{orb}$ gives rise to the asymmetric XMCD line shapes. Since the Co site is almost identical as Co$^{2+}(O_h)$ symmetry, the sum rules can be applicable for the Co XMCD spectra. The spin and orbital magnetic moments for Co$^{2+}$ sites are estimated as 1.32$\pm$0.20 and 0.63$\pm$0.09 $\mu_\mathrm{B}$, respectively, using an electron number of 7.1. The error bars mainly originate from the estimation of background in XAS which is used for the sum rule analysis. Large $m_\mathrm{orb}$ originates from the orbital degeneracy in $d^7$ electron system in the strained Co site. The multiplet structures are almost similar to the previous report by the LFM calculations along the easy-axis direction \cite{ref44,ref45}. In order to confirm the PMA, the element-specific magnetic hysteresis curves are measured at each peak in Fe and Co $L_3$ edge. As shown in Fig. 1(c), clear square shapes are detected at each photon energies due to the out-of-plane easy axis, which is almost consistent with the magnetization measurement shown in Fig. S1 \cite{ref49SM}. The square $M$-$H$ curves for all energies suggest the strong exchange coupling among each site. The opposite sign in the $M$-$H$ curves at Fe$^{3+}$ $T_d$ site and Fe$^{2+}$/Co$^{2+}$ $O_h$ sites originates from the antiferromagnetic super-exchange interaction among these sites. These suggest that the small amounts of anisotropic Co sites with large $m_\mathrm{orb}$ govern the easy axis direction of Fe sites and stabilize the PMA.

Figure 2 shows the $\bf{E}$ vector polarization dependent XAS, where the electric field $\bf{E}$ is parallel and perpendicular to the out-of-plane magnetization direction. The differential line shapes were similar to those of previously reported spectra \cite{ref44}. Because of small difference in XAS by the horizontal (parallel) and vertical (perpendicular) beams as shown in the inset of Fig. 2, XMLD intensities are also suppressed and displayed in the different scales. For Fe $L$-edge XMLD, the overlapping of three kinds of components brings complex differential line shape. We note that the integrals of the XMLD line shapes are proportional to the charge asphericity within the XMLD sum rule \cite{ref50}. For Co $L$-edge XMLD of the Co$^{2+}$ $O_h$ site, we confirmed that the integral of XMLD converges to a negative value, deducing that the sign of $Q_\mathrm{zz}$, where $Q_\mathrm{zz}$ is the diagonal tensor component of electric quadrupole moment describing the charge distribution asphericity as $Q_\mathrm{xx}+Q_\mathrm{yy}+Q_\mathrm{zz}=0$, is negative with the order of 10$^{-1}$. The result means the in-plane orbital states are strongly coupled with $\bf{E}$. This value can be also estimated from the XMCD spin sum rule considering the magnetic dipole $m_{T_z}$ term. Since the measurement geometry of 60$^\circ$ tilted from the sample surface normal cannot detect precise orthogonal direction, the oblique angle suppresses the linear dichroism intensity to $\sqrt{3}/2$. The orbital polarization of Co 3$d$ states means the pancake-type oblate charge distribution, which enhances out-of-plane component of $m_\mathrm{orb}$. Therefore, with both XMCD and XMLD combined, it can be concluded the in-plane tensile strain triggers the changes of charge distribution along in-plane direction, resulting in the large out-of-plane $m_\mathrm{orb}$ and the PMA. 

For the analysis of XMCD and XMLD spectra, we employed cluster-model calculations including the CI for Co$^{2+}$ and Fe sites in Co$_{0.2}$Fe$_{2.8}$O$_{4+\delta}$ as tetrahedral ($T_d$) TMO$_4$ and octahedral ($O_h$) TMO$_6$ clusters, modeled as a fragment of the spinel-type structures. The Hamiltonian included the electronic structure parameters of full on-site TM 3$d$–3$d$ (valence–valence) and 2$p$–3$d$ (core–valence) Coulomb interactions ($U$) and the $T_d$ or $O_h$ crystal fields (10$Dq$) in the TMs, along with the hybridization between the TM 3$d$ and O 2$p$ wave functions. The charge-transfer energy was defined as $\Delta$=$E(3d^{n+1}\underline{L})-E(3d^n)$, where $\underline{L}$ denotes a hole in a ligand $p$ orbital. The hybridization between the TM 3$d$ and O 2$p$ states was also parameterized in terms of the Slater–Koster parameters ($pd\sigma$) and ($pd\pi$), where the relation $(pd\sigma)=–1/2(pd\pi)$ was used \cite{ref51}. The parameters ($pp\sigma$) and ($pp\pi$) were always set to zero. In all cases, a Gaussian broadening was used to simulate spectral broadening. 

As shown in Fig. 3, spectral line shapes of XMCD and XMLD can be reproduced by the LFM calculations qualitatively, at least the peak positions, with three Fe states and uniformed Co state. The fitting parameters are listed in Table I. We emphasize the same adjusted parameters of $\Delta$, $U$ and $pd\sigma$ and intensity ratios of each component for XMCD and XMLD are applicable for the fitting. In comparison with the previous first-principles calculations of CFO, the values of $U=5$ eV are also plausible from the viewpoints of LFM calculations \cite{ref52,ref53,ref54,ref55}. The multiplet parameter values are almost similar to the previous report \cite{ref56}. For the Co$^{2+}$ 3$d^7$ case, the $t_{2g}$ states are split into two levels by the tetragonal distortion, and the lowest $xy$ states are occupied by one of the down spin electrons. The other electron occupies the degenerated $yz$ or $zx$ states \cite{ref35}. We determined the tetragonal distortion for the Co 3$d$ states ($D_\mathrm{tet}$) to be 0.02 eV in order to reproduce the XMCD and XMLD spectral line shapes qualitatively. However, the best fitted parameter sets shown in Table I are not sensitive to the parameters caused by the strain. The intensity ratio of Fe$^{3+}$ $T_d$, Fe$^{3+}$ $O_h$, and Fe$^{2+}$ $O_h$ is deconvoluted to be 1:1.3:0.25 in the fitting of Fe $L_3$-edge XMCD and XMLD affecting the decrease of Fe$^{2+}$ states in the inverse spinel structure. The Co $L$-edge line shapes are reproduced with only a single Co$^{2+}$ site, except the peak at 779.5 eV in XMLD. 

In order to confirm the local environment around the Co atoms, XAFS measurements with EXAFS analysis in the Co $K$-edge were performed to deduce the nearest neighbor distance through the Fourier transform. As shown in Fig. 4, the Co $K$-edge absorption spectra with EXAFS oscillatory behaviors are observed for $x=0.2$ (PMA) and $x$=0.6 (in-plane anisotropy). We note that the Fe $K$-edge XAFS cannot be separated in the Fourier transform without some assumption because of the contributions of three kinds of states. Clear XAFS oscillation is detected and plotted in the wave number $k$ in the inset of Fig. 4(a). XAFS oscillation functon $k^3{\chi}(k)$ can be fitted by FEFF8 program \cite{ref57pre} and the Fourier transformed EXAFS profile is also displayed in Fig. 4(b) considering the information up to second nearest neighbor from the Co site. By using the fitting procedure, the nearest Co-O bond length in the Co sites is 2.09 and 2.08 \AA for PMA and in-plane cases, respectively, which is almost identical with the previous study \cite{ref57}. The expansion of nearest neighbor in the PMA film is qualitatively consistent with the XRD shown in Fig. S1 \cite{ref49SM}. Because of the effect of phase shift, the peak position in EXAFS is not related to the bond length directly. Coordination number also becomes large in the case of PMA because of the distorted local environment around Co site, which corresponds to the intensity of EXAFS profile in Fig. 4(b). These suggest the tensile local distortion triggers the PMA in the Co site of strained CFO.

Considering the above results, we discuss the electronic structures of CFO. First, we discuss the suppression of Fe$^{2+}$ with increasing the Co composition. The previous report of M\"{o}ssbauer spectra for related CFO with PMA suggests the suppression of Fe$^{2+}$ \cite{ref23}, where $\sim$100 nm from the interface was probed using incident $\gamma$-rays and emitted conversion electrons. Since the probing depth region of XMCD is beneath 3 nm from the surface, finite Fe$^{2+}$ states in XMCD is located at the surface region. The previous report suggests that the Fe$^{2+}$ state is located at the surface region with inevitable lateral inhomogeneity \cite{ref57re}. 

Second, we discuss the occupation of cation sites by the charge neutrality and the number of cation elements. Using the Co concentration and charge neutrality in the inverse spinel structure of the formation of spinel structure of $AB_2$O$_4$ formula unit; Fe$^{3+(T_d)}$(Co$^{2+(O_h)}$Fe$^{2+(O_h)}$Fe$^{3+(O_h)}$)O$_4$, the following two equations are deduced for Co$_{0.2}$Fe$_{2.8}$O$_{4+\delta}$; 14$n_{\mathrm{Co}^{2+}}$=$n_{\mathrm{Fe}^{2+}}$+$n_{\mathrm{Fe}^{3+}}$, and $2n_{\mathrm{Co}^{2+}}$+2$n_{\mathrm{Fe}^{2+}}$+3$n_{\mathrm{Fe}^{3+}}$=8, where $n$ is the number of each element and the cation vacancy is implicitly incorporated as $\delta$. Since the ratio of $n_{\mathrm{Fe}^{3+}(T_d)}$, $n_{\mathrm{Fe}^{2+}(Oh)}$, and $n_{\mathrm{Fe}^{3+}(O_h)}$ is estimated from the XMCD and XMLD analyses, these four parameters are calculated to be $n_{\mathrm{Fe}^{3+}(O_h)}$=1.34, $n_{\mathrm{Fe}^{3+}(T_d)}$=1.03, $n_{\mathrm{Fe}^{2+}(O_h)}$=0.26, and $n_{\mathrm{Co}^{2+}(O_h)}$=0.19. These suggest that the existence of excess cation vacancies ($n_\mathrm{v}$) can be estimated to be 0.185 from the equation for the cation site, $n_\mathrm{v}=3-n_{\mathrm{Fe}^{3+}(O_h)}-n_{\mathrm{Fe}^{3+}(T_d)}-n_{\mathrm{Fe}^{2+}(O_h)}-n_{\mathrm{Co}^{2+}(O_h)}$. We emphasize that the cation vacancy can be estimated quantitatively from the analysis of XMCD.

Third, we discuss the relationship between strain and orbital magnetic moments. For the Co$^{2+}$ 3$d^7$ system, the ground state $^4F$ multiplet energy split by the cubic crystal field to $\Gamma_4$ state, and further effects by trigonal or tetragonal field and exchange coupling stabilize the lowest 3$d^7$ states. The splitting of $t_{2g}$ state promotes the degeneracy for spins as analogous to the 3$d^2$ system of spinel-type V$^{3+}$ compounds \cite{ref58,ref59}. Large $m_\mathrm{orb}$ is recovered by the spin-orbit coupling $\xi_{\mathrm{Co}}$ of 70 meV. Since the uniaxial anisotropy in spinel-type Co ferrite is inevitably formed from the structural and electronic configuration, the out-of-plane easy axis can originate from the tensile strain effect. In the theoretical model calculation, the PMA energy ($K$) deduced from the energy difference along the axis direction can be expressed as $K=E(100)-E(001)= -B_1\chi$, where $B_1$ and $\chi$ is the magneto-elastic coefficient and strain, respectively \cite{ref34}. Phenomenological $B_1$ of 1.4$\times$10$^8$ J/m$^3$ consists of the elastic constant and magneto-striction coefficient at room temperature for CoFe$_2$O$_4$. In the case of $x$=0.2, $B_1x$ is applied to estimate $K$. Considering the lattice constants of $a_\mathrm{\parallel}=8.42 \mathrm{\AA}$ and $a_\mathrm{\perp}=8.29 \mathrm{\AA}$ estimated from the XRD, the strain of ($a_\mathrm{\parallel}-a_\mathrm{\perp}$)/$a_0{\sim}$1.5 $\%$ was deduced using the bulk CFO value $a_0=8.38 \mathrm{\AA}$ \cite{ref32}. Therefore, the PMA energy is roughly estimated to $0.51\times10^{6}$ J/m$^3$. On the other hand, microscopic estimation of the PMA energy can be performed using anisotropic $m_\mathrm{orb}$ as $K=1/4{\xi}{\Delta}m_\mathrm{orb}$, where spin-orbit coupling constant is $\xi$ and ${\Delta}m_\mathrm{orb}$ is the difference between out-of-plane and in-plane $m_\mathrm{orb}$. Although the in-plane $m_\mathrm{orb}$ cannot be detected unless applying high magnetic field to saturate the magnetization along the hard axis, we used the value in the reference with extrapolation in low Co concentration \cite{ref44}. Assuming the value of ${\Delta}m_\mathrm{orb}$ to be 0.2 $\mu_\mathrm{B}$, the PMA energy of Co can be estimated to be $1.6\times10^6$ J/m$^3$ with the assumption of the Co composition $x$=0.2 and strained CFO volume. The value of $K$ from XMCD is overestimated compared with that from magnetization measurements, in general, by a pre-factor of 0.1 \cite{ref60}. These estimations suggest that the PMA energy can be explained mainly by the contribution of small amount of anisotropic Co sites. Furthermore, the macroscopic magnetostriction effect, which the relation between strain and $K$, can be recognized including the $m_\mathrm{orb}$ from the viewpoint of electron theory as orbital-elastic effect. The contributions from the Fe$^{3+}$ (3$d^5$) sites are negligible because of half-filled quenched orbital magnetic moments. The contribution from the Fe$^{2+}$ (3$d^6$) sites is also quenched since the strained undegenerated level is occupied. Therefore, the origin of PMA in Co$_{0.2}$Fe$_{2.8}$O$_{4+\delta}$ can be predominantly explained by the enhanced $m_\mathrm{orb}$ of 1/14 Co compositions through the strained charge distributions, which can be recognized as diluted doped magnetic oxide systems.

Finally, we discuss the origin of PMA in the spinel-type oxides from the viewpoint of tensile or compressive strain and $m_\mathrm{orb}$. The strain in CFO is derived from the tensile strain in the film, which is detected by the EXAFS analysis and confirmed by the sign of XMLD. Recent report shows that the PMA in spinel-type NiCo$_2$O$_4$ originates from the compressive strain in the film \cite{ref61,ref62} and that the enhancement of $m_\mathrm{orb}$ in XMCD measurements is not distinctive with the 2+ and 3+ mixed valence states of both Co and Ni sites \cite{ref63}. Therefore, the PMA in spinel-type oxides can be categorized by the tensile or compressive strain. Because of tensile strain, oblate-type charge distribution induces $m_\mathrm{orb}$ along out-of-plane direction, which is similar to the case of enhanced $m_\mathrm{orb}$ in Co/Pd multilayer \cite{ref66}. On the other hand, in-plane compression or out-of-plane elongation modulates the prorate-type charge distributions, which induces the quadrupole moments without enhancing $m_\mathrm{orb}$ as discussed in strained Mn$_{3-x}$Ga system \cite{ref47}.  

In summary, using XMCD and XMLD, we investigated the element-specific orbital magnetic moments and their anisotropy in Co-ferrite thin films with low Co concentration which exhibit the PMA. The origin of PMA in CFO is explained by the large $m_\mathrm{orb}$ in the Co$^{2+}$ (3$d^7$) states. The tensile lattice strain, which is deduced from EXAFS, induces the out-of-plane $m_\mathrm{orb}$ through the anisotropic in-plane charge distribution. Furthermore, a novel method to estimate the number of cation vacancy  from XMCD line shapes was also proposed. Our finding clearly reveals that the controlling strain modulates $m_\mathrm{orb}$, which opens up the material functionalities in the oxide spin-orbitronics by strain engineering, especially in the TM $d^7$ systems.

\begin{acknowledgments}
This work was partially supported by JSPS KAKENHI (Grant No. 16H06332), the Kato Science and Technology Foundation, the Izumi  Science and Technology Foundation, and the Telecommunications Advancement Foundation. Parts of the synchrotron radiation experiments were performed under the approval of the Photon Factory Program Advisory Committee, KEK (Nos. 2019G028 and 2021G069).
\end{acknowledgments}

\begin{table}[htb]
\begin{center}
\caption{The Electronic structure parameters used in the LFM cluster calculation. The units of each value is in eV.   }
\begin{tabular}{c|c|c|c|c} \hline
 & \hspace{5mm}$\Delta$\hspace{5mm} & \hspace{5mm}$U$\hspace{5mm} & \hspace{5mm}$(pd\sigma)$\hspace{5mm}  &  \hspace{5mm}10$Dq$\hspace{5mm}  \\ \hline 
Fe$^{2+}$ ($O_h$)  &  6.5  &  6.0  &  1.2  &  0.9  \\
Fe$^{3+}$ ($O_h$)  &  0.5  &  6.0  &  1.2  &  0.9  \\
Fe$^{3+}$ ($T_d$)  &  4.5  &  6.0  &  2.0  &  -0.5  \\
Co$^{2+}$ ($O_h$)  &  6.5  &  6.0  &  1.3 &  0.5  \\ \hline 

\end{tabular}
\end{center}
\end{table}

\begin{figure}
[ptb]
\begin{center}
\includegraphics[
width=5.5in
]%
{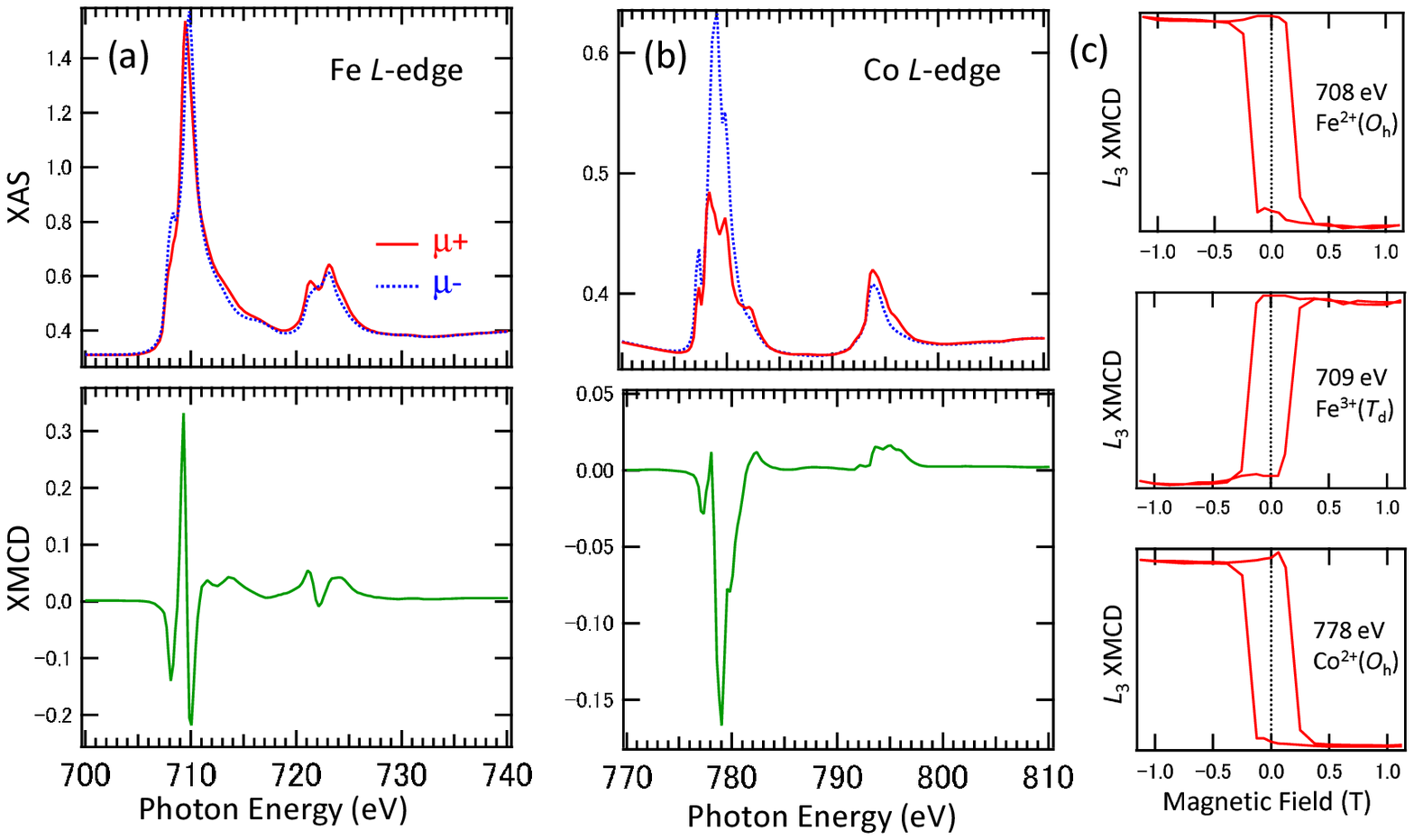}%
\end{center}
\caption{XAS and XMCD of Co$_{0.2}$Fe$_{2.8}$O$_{4+{\delta}}$ film. $\mu$+ and $\mu-$ in XAS denote the magnetic field direction along the incident photon beams at normal incident geometry. Difference of ${\mu^+}-{\mu^-}$ is defined as XMCD spectra for (a) Fe $L$-edge and (b) Co $L$-edge. Both XAS and XMCD intensities are normalized by photon flux. (c) Magnetic-field dependence at each fixed photon energy; 708, 709, and 778 eV corresponding to Fe$^{2+}$ $(O_h)$, Fe$^{3+}$ $(T_d)$, and Co$^{2+}$ $(O_h)$ peaks, respectively.}
\end{figure}

\begin{figure}
[ptb]
\begin{center}
\includegraphics[
width=5.5in
]%
{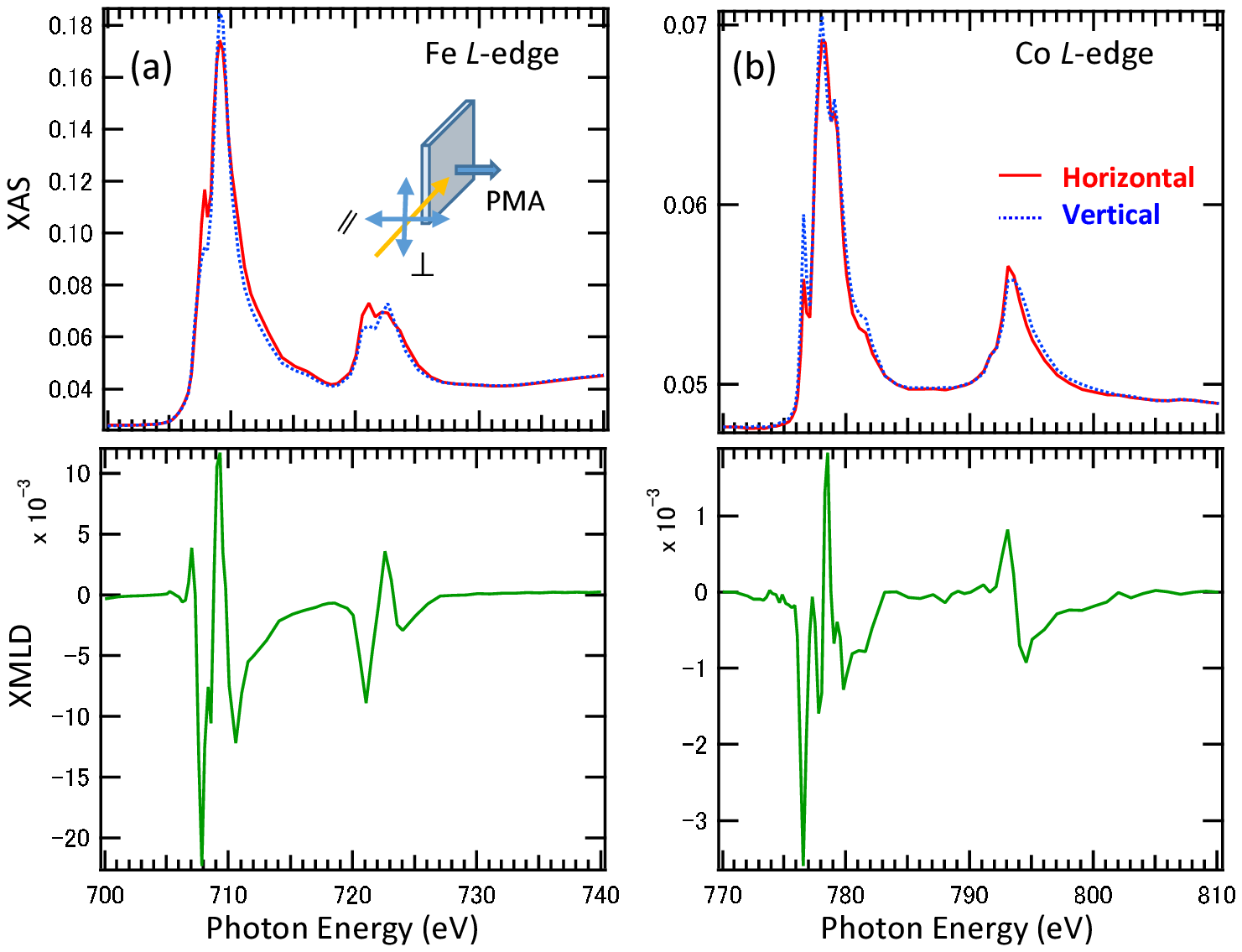}%
\end{center}
\caption{XAS and XMLD of Co$_{0.2}$Fe$_{2.8}$O$_{4+\delta}$ film of (a) Fe and (b) Co $L$-edges. Spectra were taken at the grazing incident setup where $\bf{E}$ of the incident beam and direction of magnetization $\bf{M}$ were parallel (horizontal) and perpendicular (vertical), respectively. The inset displays an illustration of the XMLD measurement geometry. }
\end{figure}

\begin{figure}
[ptb]
\begin{center}
\includegraphics[
width=5.5in
]%
{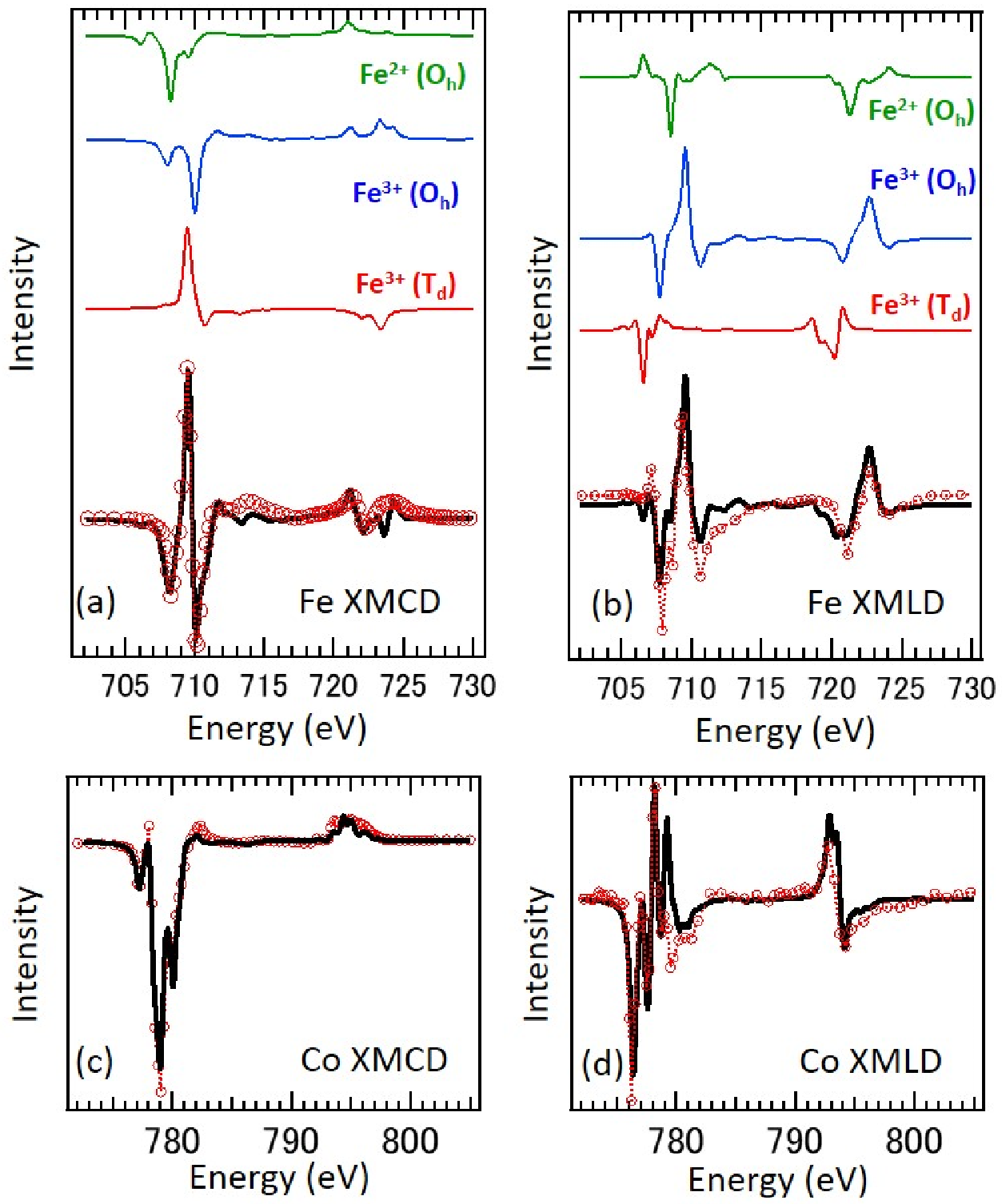}%
\end{center}
\caption{Calculated XMCD and XMLD spectra for (a,b) Fe and (c,d) Co $L$-edges. For Fe $L$-edge, three components of Fe$^{2+}$ $(O_h)$, Fe$^{3+}$ $(O_h)$, and Fe$^{3+}$ $(T_d)$ sites are also shown. Total spectra are shown in black solid line. Open dotted curves are experimental XMCD and XMLD shown in Figs. 1 and 2.  }
\end{figure}

\begin{figure}
[ptb]
\begin{center}
\includegraphics[
width=5.5in
]%
%{fig4AuFe_re.eps}%
{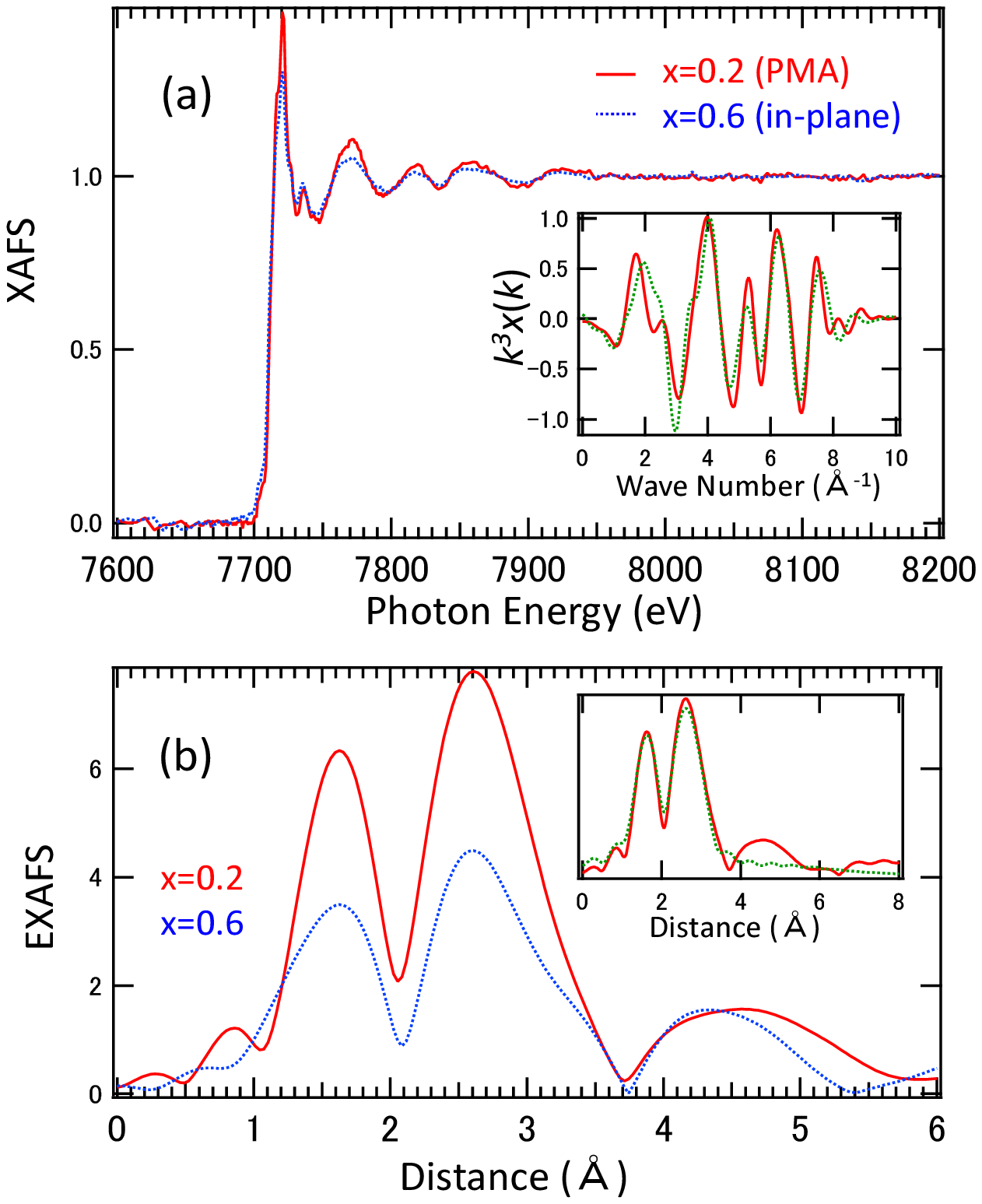}%
\end{center}
\newpage
\caption{XAFS and EXAFS of Co$_{0.2}$Fe$_{2.8}$O$_{4+\delta}$ (PMA) and Co$_{0.6}$Fe$_{2.4}$O$_{4+\delta}$ (in-plane anisotropy). (a) Co $K$-edge XAFS. Inset shows corresponding $k$ oscillation in $k^3\chi(k)$ for $x=0.2$. Dot curve shows the fitting result. (b) EXAFS profile after the Fourier transform. Inset shows the fitting result in $x=0.2$.  }   
\end{figure}

\end{document}